\documentclass[aps,prl,preprint,groupedaddress,showpacs,showkeys]{revtex4}
\usepackage{graphicx}
\usepackage[figuresleft]{rotating}
\usepackage{makecell,rotating}
\usepackage{longtable}

\begin{document}


\title{An alternative indicator of annihilated electrons in atoms: Rahm's electronegativity scale}
\author{Yu Wu}
\author{Xiaoguang Ma}
\email{hsiaoguangma@ldu.edu.cn}
\author{Fang Yuan}
\author{Jipeng Sui}
\author{Meishan Wang}
\author{Chuanlu Yang}
\affiliation{School of Physics and Optoelectronic Engineering, Ludong University, Yantai, Shandong 264025, People's Republic of China}
\date{\today}

\begin{abstract}
This paper presents a new explanation of the width of gamma-ray spectra based on Rahm's electronegativity scale. This quantitatively proves, for the first time, that positrophilic electrons in the positron-electron annihilation process are exactly the valence electrons. This suggests the replacement of Full Width at Half Maximum (FWHM) of the gamma-ray spectra with the newly defined physical quantity Average Doppler Shift (ADS). Both FWHM and ADS of the gamma-ray spectra in light elements agree well with the corresponding Rahm's electronegativity values, respectively. Further, ADS provides strong evidence in favor of Rahm's electronegativity scale. It is expected that this will be useful in understanding the mechanism of the positron-electron annihilation process.
\end{abstract}

\pacs{78.70.Bj, 82.30.Gg, 36.10.Dr}
\keywords{positron-electron annihilation; positrophilic electrons; electronegativity; average Doppler shift}

\maketitle

\section{Introduction\label{}}
The gamma-ray spectra for low-energy positron annihilation in molecules have been studied extensively over the several past decades\cite{1,2,3,4,5}. However, understanding of low-energy positrons and their behavior in molecules is still incomplete compared to more familiar electron problems\cite{6,7,8}. Most of the experimental measurements exhibit good agreement with the theoretical gamma-ray spectra of the valence electrons in molecules\cite{7,8,9,10}. However, next to no research has been conducted to quantitatively establish the reason why valence electrons dominate the positron-electron annihilation process.

Moreover, the width FWHM of the gamma-ray spectra is only an experimental spectrum analysis parameter, which has almost no physical interpretation. In other words, it is just a technical term, and does not correspond to a physical quantity. This makes it difficult to understand the mechanism of the positron-electron annihilation process. For instance, the reason behind the empirically observed higher annihilation rate compared to the theoretical prediction is one of many issues that eludes explanation. Some explanations and suggestions based on the comparison between theoretical calculations and experimental measurements have been given. However, the physical meaning of the width of the gamma-ray spectra remains elusive.

Recently, M. Rahm et. al. introduced a new electronegativity scale\cite{11}, which provides some hints to explain the dominance of valence electrons in the positron-electron annihilation process. Electronegativity assigns a measure to atoms (actually nuclear or positrons) based on their proclivity to attract electrons in a molecule. It is an important physical quantity in chemical reactions. It can naturally be used to find positrophilic electrons in molecules.

The gamma-ray spectra are generally represented by the Doppler shift parameter. The Doppler shift, i.e., the width (usually using the full width at half maximum, FWHM) of gamma-ray spectra represents the probability of the existence of positron-electron pairs at certain energies. The integration of the gamma-ray spectra produces the annihilation rate, i.e., the total number of electrons annihilated with an incoming positron\cite{10}. The maximum theoretical annihilation rate is taken to be the number of electrons in a molecule in general theories\cite{8}. However, as mentioned previously, experimental measurements indicate that the annihilation rate is usually considerably larger than the number of electrons in a molecule\cite{9}. Moreover, only the number of valence electrons has been usually considered in translating the annihilation process.

In order to fundamentally understand the mechanism of the positron-electron annihilation process, identification of a new physical quantity to represent the gamma-ray spectra and to explain the positrophilic electrons in the positron-electron annihilation process is necessary. The present study attempts to identify such an alternative physical quantity and explain the dominance of valence electrons in the positron-electron annihilation process by using Rahm's electronegativity scale.

\section{The physical parameters of gamma-ray spectra\label{}}
During the annihilation process, the momentum $\vec{p}$ of an emitted photon is equal to the momentum of the parent positron-electron pair, i.e., $\vec{p}=\vec{k^-}+\vec{k^+}$. Therefore, the momentum distribution of photons can be obtained via Fourier transform from the positron-electron pair wavefunction\cite{10}
\begin{equation}
A(\vec{p})=\int{\Phi^-({\vec{r}})\Phi^+(\vec{r})e^{-i\vec{p}\cdot\vec{r}}d\vec{r}},
\end{equation}
where $\Phi^-({\vec{r}})$ and $\Phi^+(\vec{r})$ denote the wavefunctions for electrons and positrons in the real space, respectively.

The momentum of the electron-positron pair is rotationally averaged in gas or liquid experiments. Hence, in order to compare our results with these experimental measurements, the theoretical momentum distribution must be spherically averaged. The radial distribution function in the momentum space can be defined by
\begin{equation}
D(p)=\int_0^\pi{d\theta}\int_0^{2\pi}{d\phi{P^2}\sin{\theta}|A(\vec{p})|^2},
\end{equation}
where $P,\theta,\phi$ denote spherical coordinates. Hence, the theoretical spherically averaged momentum distribution is given
\begin{equation}
\sigma(p)=\frac{D(p)}{4\pi{p}^2}.
\end{equation}
which gives the mean probability to encounter the electron-positron pair on the surface with the momentum $|P|$.

Following this, the gamma-ray spectra during the annihilation process is Doppler shifted in energy due to the longitudinal momentum component of the positron-electron pair\cite{8}. Hence, integration must be performed over the plane perpendicular to $p$ in order to obtain the total probability density corresponding to the momentum $p=2\epsilon/c$. Then, the gamma-ray spectra for the positron-electron pair are given by
\begin{equation}
\Omega(\epsilon)=\frac{1}{c}\int_{2\epsilon/c}^{\infty}{\sigma(p)pdp}.
\end{equation}
The Doppler shift from the center ($mc^2=511$ keV) is given by $\epsilon$. For low-energy positrons, the gamma-ray spectra are dominated by the bound electrons in molecules. The Doppler shift $\epsilon$ is directly related to the bonding energy $\epsilon_n$ of the annihilated bound electrons $\epsilon{\propto}\sqrt{\epsilon_n}$\cite{10}. In other words, the Doppler shift ($\epsilon$) is determined by the bonding energy of the annihilated bound electrons. If we integrate the gamma-ray spectra, the annihilation rate is given by
\begin{equation}
Z_{eff}=\int{\Omega(\epsilon)d\epsilon}.
\end{equation}
Then, the average binding energy (Doppler shift, actually) of the annihilation electrons will be given by
\begin{equation}
\bar{\epsilon}=\frac{\int{\Omega(\epsilon)}{\epsilon}d\epsilon}{\int{\Omega(\epsilon)d\epsilon}}.
\end{equation}
The gamma-ray spectra carry information about the electron momentum distribution in the bound state orbitals\cite{1}. In the present study, $\bar{\epsilon}$ is named the Average Doppler Shift (ADS). The Doppler shift of the gamma-ray spectra is proportional to the square root of the absolute value of the binding energy of the annihilated bound electrons\cite{10}. Hence, ADS of gamma-ray spectra represents the average bonding energy of the positron-electron pairs.

ADS is a more general measure than FWHM of gamma-ray spectra in terms of representing the gamma-ray spectra, because FWHM is merely an analysis parameter of the spectrum shape with only one peak. Further, FWHM is merely a special point while ADS is obtained by averaging over all the points of the gamma-ray spectra. A general gamma-ray spectrum comprises more than one peak and shoulder. Hence, FWHM is not applicable in these complicated situations. However, ADS is determined by taking the entire data of gamma-ray spectra into account, and not merely one value at half maximum. Hence, ADS is capable of representing all characteristics of gamma-ray spectra.

Moreover, for a low-energy positron annihilation process, the ADS of valence electrons in a molecule also has the identical definition and agrees well with Rahm's electronegativity scale\cite{11}. Therefore, in order to understand the mechanism of the positron-electron annihilation process, the experimental analysis parameters, such as FWHM, should be replaced by the physical quantity ADS. Eq.(6) remains the same and is consistent with Rahm's electronegativity scale\cite{11}
\begin{equation}
\bar{\chi}=\frac{\sum_{i=1}^{N}n_i\epsilon_i}{\sum_{i=1}^{N}n_i}.
\end{equation}
with the summation running over the entire set of valence electrons in a molecule. Eq.(6) and Eq.(7) both present very natural definitions that measure the ability to attract positrons. Comparing them with ADS of gamma-ray spectra and Rahm's electronegativity value, one can quantitatively prove the dominance of the valence electrons during the annihilation process.
\begin{figure}
\centering
\includegraphics[width=1.0\textwidth, angle=0]{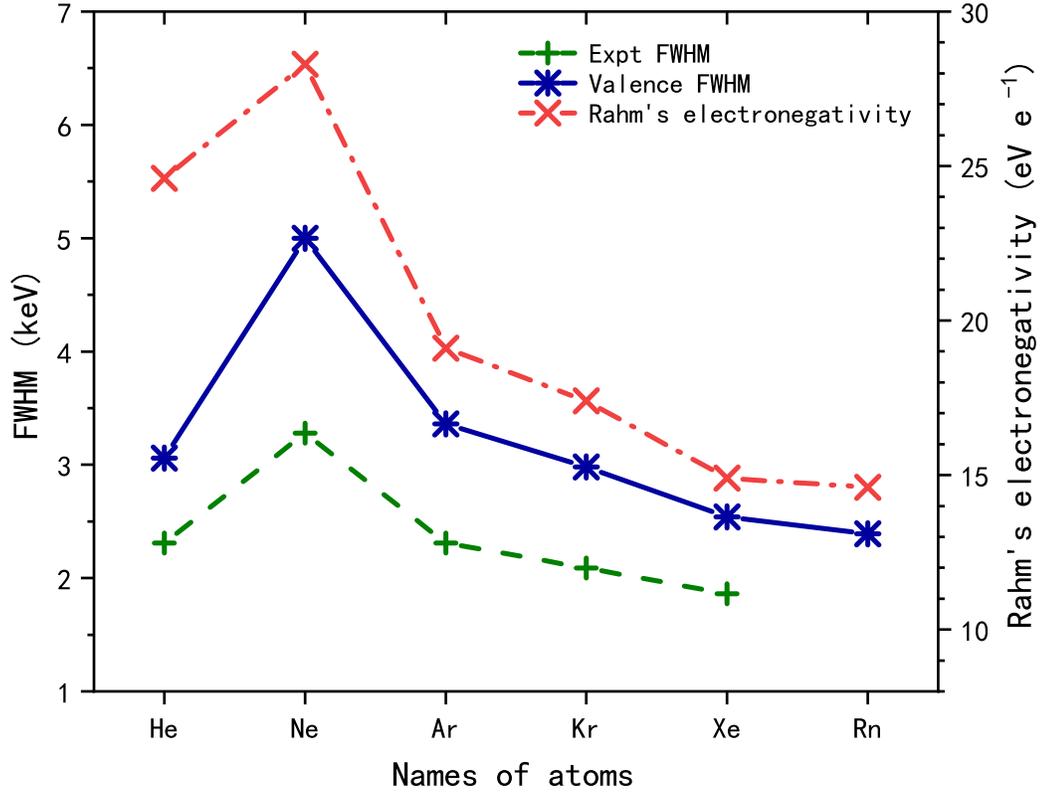}
\caption{\label{HeFig} The experimental and theoretical FWHM of gamma-ray spectra of noble gas atoms compared with Rahm's electronegativity.}
\end{figure}

\section{Application and discussion\label{}}

\begin{figure}
\centering
\includegraphics[width=1.0\textwidth, angle=0]{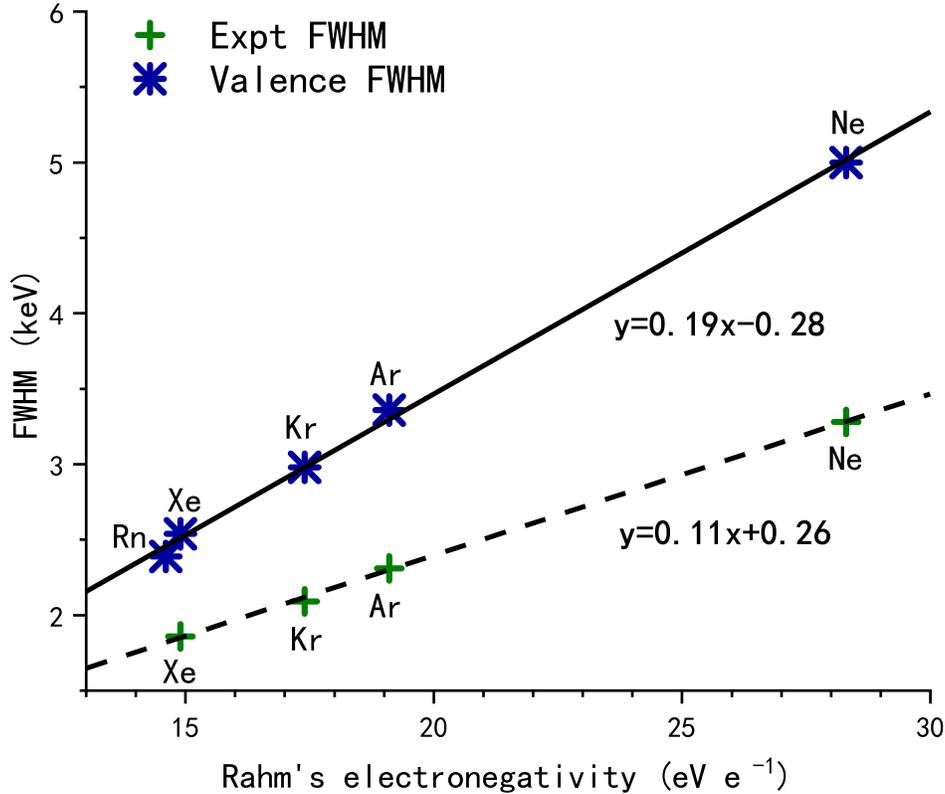}
\caption{\label{NeFig} The experimental and theoretical FWHM of gamma-ray spectra of noble gas atoms as functions of their electronegativities.}
\end{figure}
The recently developed Rahm's electronegativity scale\cite{11} provides significant hints to explain the physical meaning of the gamma-ray spectra. In most of the previous studies, the characteristics of gamma-ray spectra have always been represented by the width FWHM of the profiles\cite{1,2,3,4}. However, FWHM represents only one point in the gamma-ray spectra which lacks a direct physical interpretation. ADS takes into account all the points in gamma-ray spectra and incorporates the entirety of the information in itself. In the benchmark of the experimental studies of gamma-ray spectra\cite{9}, the C. M. Surko group presented recent  accurate measurements of the gamma-ray spectra for several molecules. In order to explain their measurements, one-Gaussian function was fitted to their measurements, leading to the loss of certain structures, such as the shoulders of the profiles. They also mentioned that their presentation of experimental data was merely analytical and quantitative, acknowledging that the physical meanings of the fitting parameters remain unknown.\cite{9} Their measurements were precise enough to study the line shape of the spectra and not just the width\cite{1}. Moreover, most experimental gamma-ray spectra exhibit some shoulders, and are not exactly smooth to fit. The presence of more than one shoulder in a gamma-ray spectrum necessitates multiple Gaussian functions to fit the spectrum approximately\cite{9}. Hence, a reasonable physical interpretation of the gamma-ray spectra seems necessary.

\begin{figure}
\centering
\includegraphics[width=1.0\textwidth, angle=0]{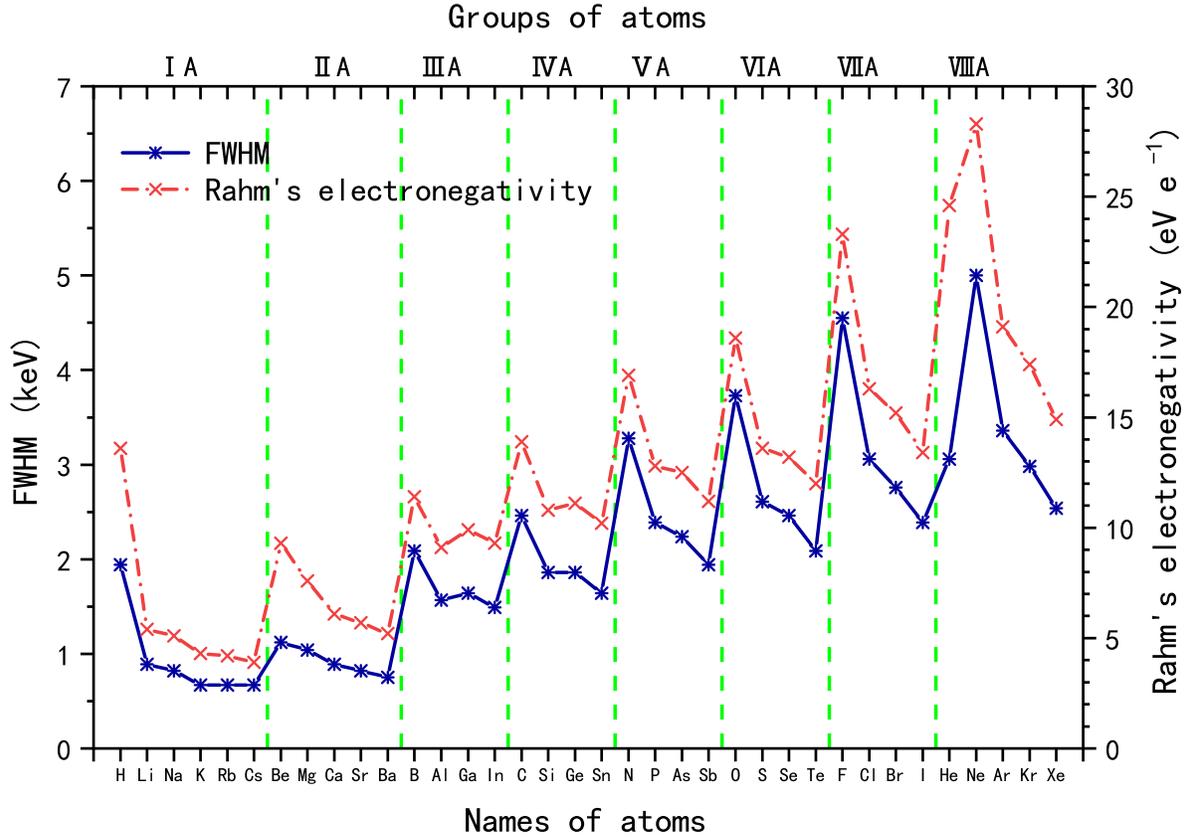}
\caption{\label{H2Fig} The Comparison between FWHM of gamma-ray spectra and electronegativity of all the 36 atoms.}
\end{figure}
As depicted in Fig.1 of reference\cite{3} and Fig.2 of reference\cite{4}, the experimentally obtained widths of gamma-ray spectra for several molecules have been compared with the corresponding theoretical predictions. The comparison reveals that the two values agree well for valence electrons. This induces the conclusion that valence electrons in molecules dominate the annihilation process. However, the reason behind this is unclear. A physical meaning of the width of gamma-ray spectra is also elusive. The benchmark measurements are precise enough to study the line shape of the spectra, and not just their widths\cite{9}. Hence, a new method to translate the spectra needs to be developed. As depicted in Fig.1, in addition to the comparison of the experimental and theoretical width, we also compared the width with Rahm's electronegativity. Electronegativity assigns a measure to the propensity of atoms in molecules to attract electrons\cite{11}. In the present work, the exact definition of the electronegativity of an atom is alternatively taken to be its ability to attract positrons.

\begin{figure}
\centering
\includegraphics[width=1.0\textwidth, angle=0]{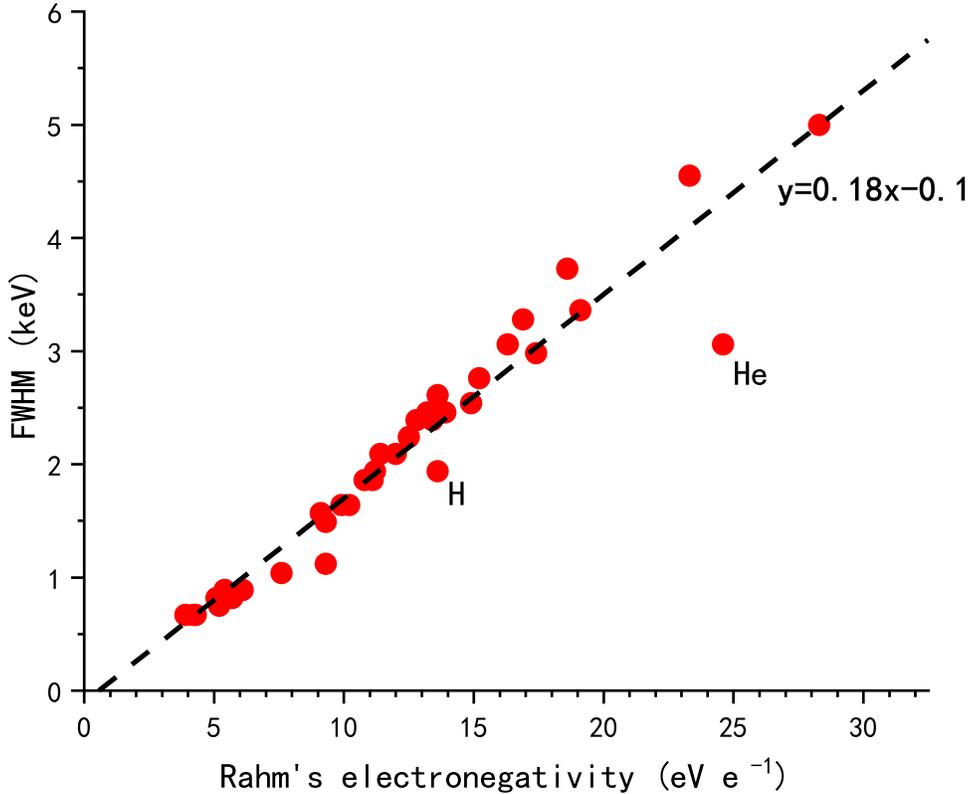}
\caption{\label{CH4Fig}FWHM of gamma-ray spectra of 36 atoms as functions of their electronegativities.}
\end{figure}
We find that the electronegativity of a noble gas atom exhibits the same variation rule with respect to the widths of the gamma-ray spectra. According to Rahm's electronegativity scale, electronegativity is defined to be the average valence electron binding energy in atoms in Eq.(7). On the one hand, good agreement of electronegativity with the experimental measurement implies the dominance of valence electrons during the annihilation process. On the other hand, the width of gamma-ray spectra might be related to electronegativity, or the average valence electron binding energy in atoms.

\begin{figure}
\centering
\includegraphics[width=1.0\textwidth, angle=0]{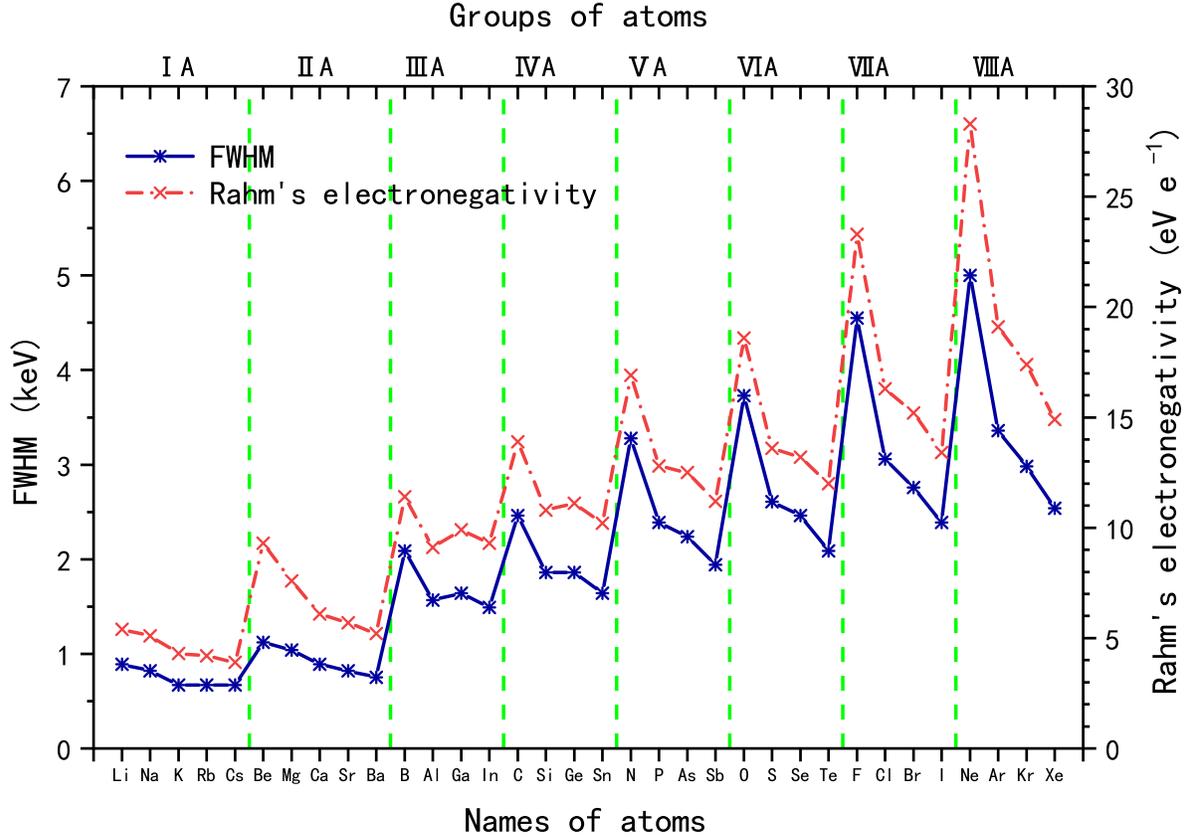}
\caption{\label{H2Fig} The Comparison between FWHM of gamma-ray spectra and electronegativities of 34 atoms without helium and hydrogen.}
\end{figure}
As depicted in Fig.2, both the theoretical and the experimental widths are observed to be linearly related to the electronegativity of the corresponding atoms. This strong linear correlation of electronegativity with the widths of the gamma-ray spectra hints at the presence of positrophilic electrons in molecules. It is known that Rahm's electronegativity scale is defined by averaging the ground state valence electron bonding energies. The agreement of the experimental widths of gamma-ray spectra with Rahm's electronegativity verifies that the valence electrons dominate the annihilation process. The present calculated width for the valence electrons has been depicted in Fig.2 and it also agrees well with Rahm's electronegativity. This probably indicates that the width of gamma-ray spectra shares a strong relationship with electronegativity. Further, whether or not the width represents the average valence electron binding energy is also a relevant question. Only helium exhibits behavior different from that of other noble gas atoms both with respect to width and electronegativity. This is explored later in the present work.

\begin{figure}
\centering
\includegraphics[width=1.0\textwidth, angle=0]{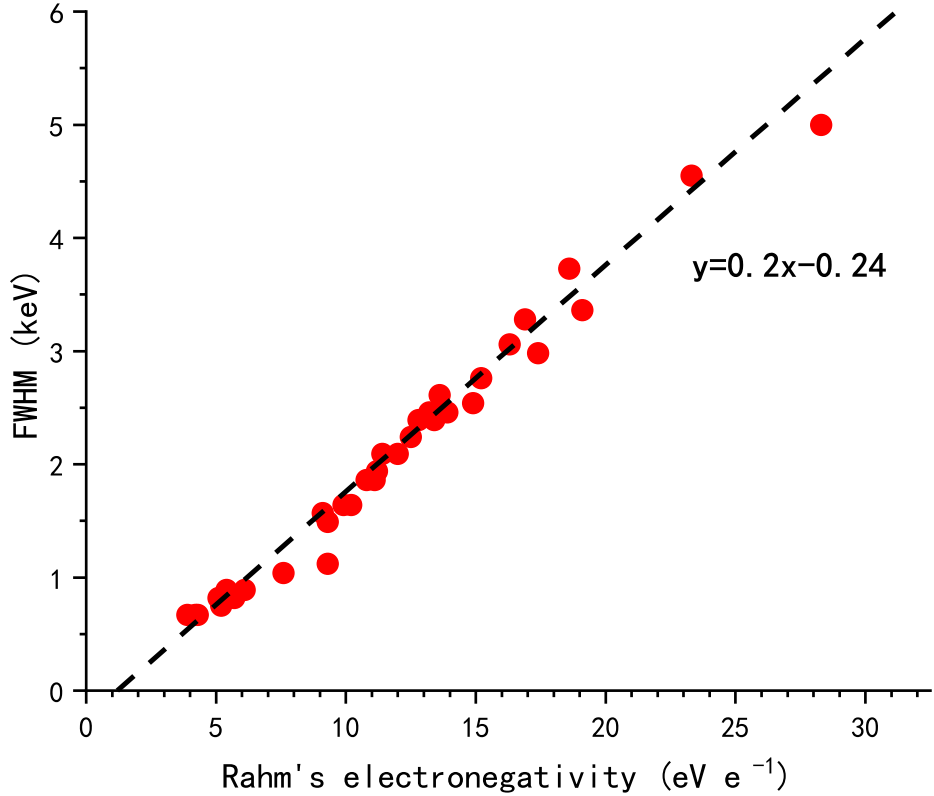}
\caption{\label{H2Fig}FWHM of gamma-ray spectra of 34 atoms as functions of their electronegativities except helium and hydrogen.}
\end{figure}
In order to identify the physical meaning of the width of a gamma-ray spectrum and the relationship it shares with electronegativity, we studied the gamma-ray spectra of 36 atoms corresponding to 1-56 elements of group A in the periodic table, as depicted in Fig.3 and Fig.4. The valence electrons of these atoms have been defined by Rahm\cite{11}. Heavy and group B atoms were not considered due to the difficulty of obtaining accurate wavefunctions for them. In Fig.3, the widths of the gamma-ray spectra of 36 light atoms have been compared with their corresponding Rahm's electronegativities. Within the same group, the variation in the width of gamma-ray spectra is observed to be completely consistent with the variation in electronegativity of these atoms. As the widths and the electronegativities are obtained from the same valence electrons, the notion that the width of the gamma-ray spectra is related with Rahm's electronegativity of atoms is further strengthened. The width of gamma-ray spectra and Rahm's electronegativity are observed to share an obvious linear correlation, as depicted in Fig.4. The electronegativity of all A group elements were basically linearly related with their corresponding widths, except for helium. Thus, Rahm's electronegativity can be considered to be an alternative indicator of the annihilated electrons in molecules.

\begin{figure}
\centering
\includegraphics[width=1.0\textwidth, angle=0]{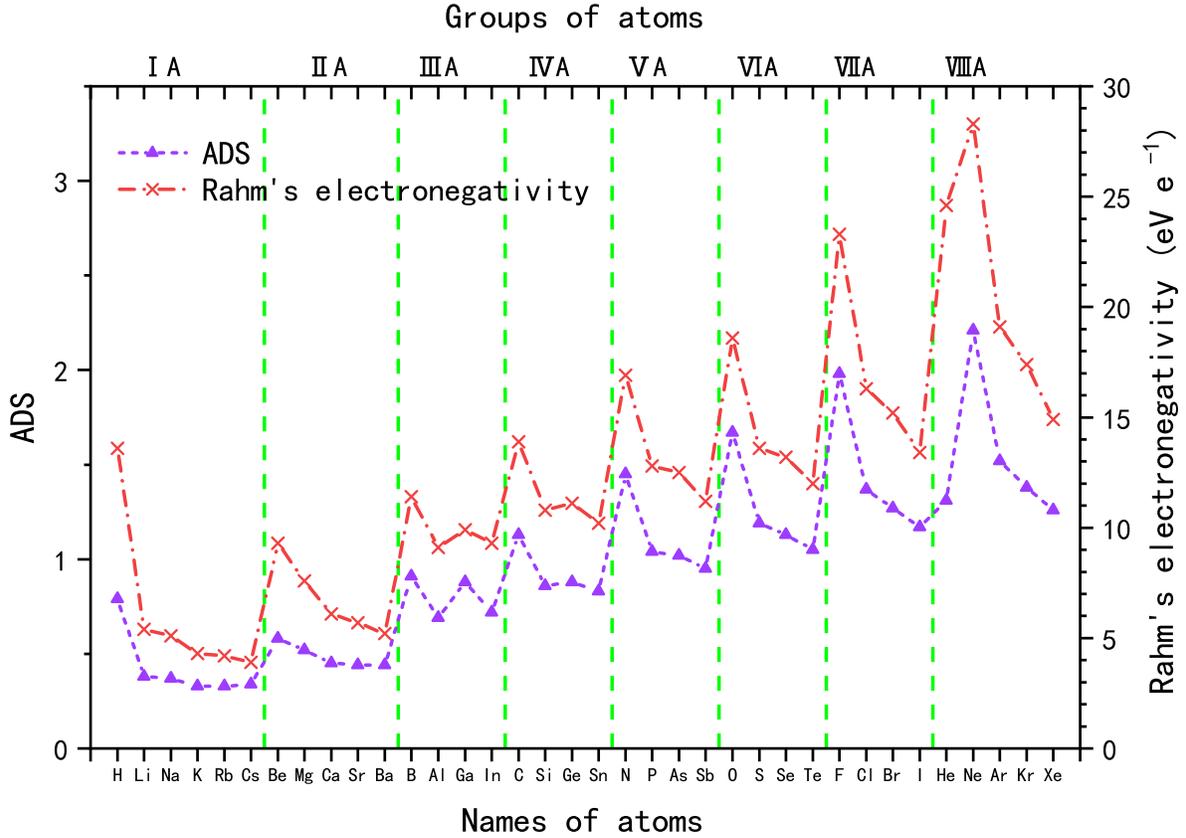}
\caption{\label{H2Fig} The Comparison between ADS of gamma-ray spectra and electronegativity of 36 atoms. }
\end{figure}
Fig.5 compares the electronegativities and widths of gamma spectra for 34 atoms except hydrogen and helium. The linear relationship only exhibited slight differences over the 36 atoms. Based on the aforementioned analysis, it can be concluded that the widths of gamma-ray spectra produced by the annihilation of positrons in atoms are closely related to the electronegativities of the corresponding atoms. This relationship can be used to obtain a physical interpretation of the width of a gamma-ray spectrum. Further obvious evidence can be found in Fig.6. The width of gamma-ray spectra might represent the ability of valence electrons to attract a positron. Hence, electronegativity was observed to be about five times of the FWHM of the gamma-ray spectra. As mentioned previously, FWHM is an ideal parameter used to characterize the shape of the spectrum, and lacks any physical meaning. Eq.(6) associates real physical meaning to the real quantity ADS. The prospective relationship between ADS and electronegativity is also explored.

\begin{figure}
\centering
\includegraphics[width=1.0\textwidth, angle=0]{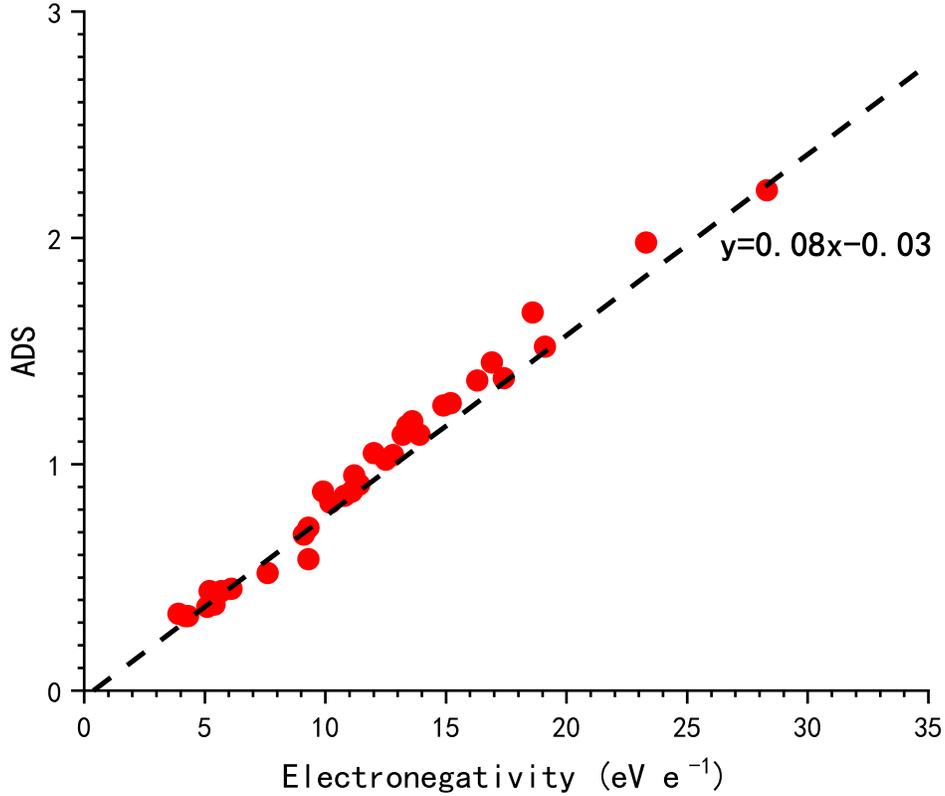}
\caption{\label{H2Fig} ADS of gamma-ray spectra of 34 atoms as functions of their electronegativities except helium and hydrogen.}
\end{figure}
In Fig.7, we present the ADS of gamma-ray spectra for 36 atoms. As is apparent, ADS exhibits identical variations as electronegativity corresponding to the group A elements considered. From group IA to VIIIA, both electronegativity of atoms and ADS of the corresponding atomic gamma-ray spectra were observed to increase slowly. The first elements in each group exhibited maximum electronegativity and ADS, except for helium in group VIIIA in which neon exhibited the maximum. On the other hand, the maximum in each group was observed to increase from IA to VIIIA, except for hydrogen in group IA. The electronegativity and ADS, and even FWHM as presented in Fig.3, of hydrogen were larger than the largest values in groups II and III.

This implies that ADS is capable of representing the gamma-ray spectra comparably to the FWHM parameters. Fig.8 depicts a strong linear relationship between ADS electronegativity. Meanwhile, the similarity in behavior with electronegativity demonstrates the ADS can represent the ability of an atom to attract a positron. ADS encodes the ability to annihilate an electron with positron. As ADS arises from the integration of the entire gamma-ray spectrum, it is more general than FWHM. Furthermore, based on Eq.(6) ADS represents the average Doppler shifts of the gamma-ray spectra, and thus inherently possesses obvious physical meaning.

\begin{figure}
\centering
\includegraphics[width=1.0\textwidth, angle=0]{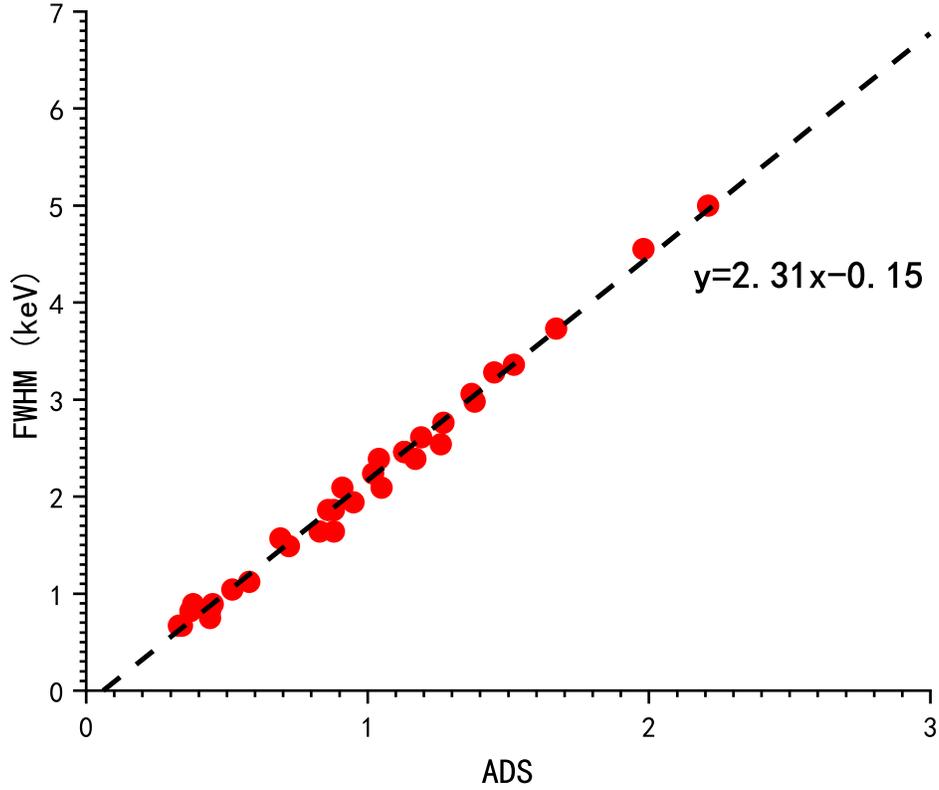}
\caption{\label{H2Fig} ADS of gamma-ray spectra of 34 atoms as functions of their FWHMs, except helium and hydrogen.}
\end{figure}
Like the widths of the gamma-ray spectra, ADS can be taken to be an indicator of positrophilic electrons, as illustrated in Fig.8. A strong linear relationship was also found between ADS and electronegativity. In other words, just like electronegativity, the average Doppler shift of a gamma-ray spectrum represents the ability of the corresponding atom to attract the incoming positrons. The ratio between ADS and FWHM is approximately 0.43, as given in Fig.9. The ADS of gamma-ray spectra can certainly be used to describe the gamma-ray spectra of positron-electron annihilation processes.

\section{Conclusoin\label{}}
In the present study, the physical quantity Average Doppler Shift (ADS) was introduced to represent the characteristics of gamma-ray spectra instead of the analysis parameter FWHM. The ADS of the gamma-ray spectra exhibit similar physical meaning as the average energy of the positron-electron pairs. The ADS of the gamma-ray spectra for the elements of group A below the atomic number 56 were observed to agree well with the corresponding Rahm's electronegativity values. Using ADS of gamma-ray spectra, the positrophilic electrons in the positron-electron annihilation process were, for the first time, quantitatively proven to be the valence electrons. Based on Rahm's electronegativity scale, it is concluded that the present ADS is useful in understanding the mechanism of the positron-electron annihilation process rather than the width (FWHM) of the gamma-ray spectra.

\textit{Acknowledgement- }This work is supported by the National Natural Science Foundation of China under grants No. 11674145 and Taishan Scholars Project of Shandong province (Project No. ts2015110055).


\begin{thebibliography}{00}
\bibitem{1} Swann A R and Gribakin G F 2019 \textit{ Phys. Rev. Lett. } {\bf39} 1647.
\bibitem{2} Swann A R and Gribakin G F 2018 \textit{ J. Chem. Phys.} {\bf149} 2018.
\bibitem{3} Ma X G, Wang M S, Zhu Y H and Yang C L 2017 \textit{ Phys. Rev. A } {\bf95} 036702.
\bibitem{4} Ma X G, Wang M S, Zhu Y H, Liu Y, Yang C L and Wang D H 2016 \textit{ Phys. Rev. A } {\bf94} 052709.
\bibitem{5} Ma X G, Zhu Y H, Liu Y 2015 \textit{ Phys. Lett. A} {\bf379} 2306.
\bibitem{6} Green D G and Gribakin G F 2017 \textit{ Phys. Rev. A } {\bf95} 036701.
\bibitem{7} Ma X G, Wang L Z and Yang C L 2014 \textit{ Phys. Lett. A } {\bf378} 1126.
\bibitem{8} Gribakin G F, Young J A and Surko C M 2010 \textit{ Rev. Mod. Phys.} {\bf82} 2557.
\bibitem{9} Iwata K, Greaves R G and Surko C M 1997 \textit{ Phys. Rev. A } {\bf55} 3586.
\bibitem{10} Dunlop L J M  and Gribakin G F  2006 \textit{ J.Phys B } {\bf39} 1647.
\bibitem{11} Martin R, Tao Z, and Roald H 2019 \textit{ J. Am. Chem. Soc. } {\bf141} 342-351.
\end{thebibliography}

\end{document}